# Modeling and Simulation of Scalable Cloud Computing Environments and the CloudSim Toolkit: Challenges and Opportunities

Rajkumar Buyya[1], Rajiv Ranjan[2] and Rodrigo N. Calheiros[1,3]

[1] **Gri**d Computing and **D**istributed **S**ystems (GRIDS) Laboratory
Department of Computer Science and Software Engineering
The University of Melbourne, Australia

[2] Department of Computer Science and Engineering
The University of New South Wales, Sydney, Australia

[3]Pontifical Catholic University of Rio Grande do Sul
Porto Alegre, Brazil
Email: {raj, rodrigoc}@csse.unimelb.edu.au, rajiv@unsw.edu.au

## Abstract

*Cloud computing aims to power the next generation data centers and enables application service providers to lease data center capabilities for deploying applications depending on user QoS (Quality of Service) requirements. Cloud applications have different composition, configuration, and deployment requirements. Quantifying the performance of resource allocation policies and application scheduling algorithms at finer details in Cloud computing environments for different application and service models under varying load, energy performance (power consumption, heat dissipation), and system size is a challenging problem to tackle. To simplify this process, in this paper we propose CloudSim: an extensible simulation toolkit that enables modelling and simulation of Cloud computing environments. The CloudSim toolkit supports modelling and creation of one or more virtual machines (VMs) on a simulated node of a Data Center, jobs, and their mapping to suitable VMs. It also allows simulation of multiple Data Centers to enable a study on federation and associated policies for migration of VMs for reliability and automatic scaling of applications.*

## 1. Introduction

Cloud computing delivers infrastructure, platform, and software as services, which are made available as subscription-based services in a pay-as-you-go model to consumers. These services in industry are respectively referred to as Infrastructure as a Service (IaaS), Platform as a Service (PaaS), and Software as a Service (SaaS). The importance of these services is highlighted in a recent report from Berkeley as: "Cloud computing, the long-held dream of computing as a utility, has the potential to transform a large part of the IT industry, making software even more attractive as a service" [11].

Clouds [10] aim to power the next generation data centers by exposing them as a network of virtual services (hardware, database, user-interface, application logic) so that users are able to access and deploy applications from anywhere in the world on demand at competitive costs depending on users QoS (Quality of Service) requirements [1]. Developers with innovative ideas for new Internet services are no longer required to make large capital outlays in the hardware and software infrastructures to deploy their services or human expense to operate it [11]. It offers significant benefit to IT companies by freeing them from the low level task of setting up basic hardware and software infrastructures and thus enabling more focus on innovation and creation of business values.

Some of the traditional and emerging Cloud-based applications include social networking, web hosting, content delivery, and real time instrumented data processing. Each of these application types has different composition, configuration, and deployment requirements. Quantifying the performance of scheduling and allocation policies in a real Cloud environment for different application and service models under different conditions is extremely challenging because: (i) Clouds exhibit varying demand, supply patterns, and system size; and (ii) users have heterogenous and competing QoS requirements. The use of real infrastructures such as Amazon EC2, limits the experiments to the scale of the infrastructure, and makes the reproduction of results an extremely difficult undertaking. The main reason for this being the conditions prevailing in the Internet-based environments are beyond the control of developers of resource allocation and application scheduling algorithms.

An alternative is the utilization of simulation tools that open the possibility of evaluating the hypothesis prior to software development in an environment where one can reproduce tests. Specifically in the case of Cloud computing, where access to the infrastructure incurs payments in real currency, simulation-based approaches offer significant benefits to Cloud customers by allowing them to: (i) test their services in repeatable and controllable environment free of cost; and (ii) tune the performance bottlenecks before deploying on real Clouds. At the provider side, simulation environments allow evaluation of different kinds of resource leasing scenarios under varying load and pricing distributions. Such studies could aid providers in optimizing the resource access cost with focus on improving profits. In the absence of such simulation platforms, Cloud customers and providers have to rely either on theoretical and imprecise evaluations, or on try-and-error approaches that lead to inefficient service performance and revenue generation.

Considering that none of the current distributed system simulators [4][7][9] offer the environment that can be directly used by the Cloud computing community, we propose CloudSim: a new, generalized, and extensible simulation framework that enables seamless modeling, simulation, and experimentation of emerging Cloud computing infrastructures and application services. By using CloudSim, researchers and industry-based developers can focus on specific system design issues that they want to investigate, without getting concerned about the low level details related to Cloud-based infrastructures and services.

CloudSim offers the following novel features: (i) support for modeling and simulation of large scale Cloud computing infrastructure, including data centers on a single physical computing node; and (ii) a self-contained platform for modeling data centers, service brokers, scheduling, and allocations policies. Among the unique features of CloudSim, there are: (i) availability of virtualization engine, which aids in creation and management of multiple, independent, and co-hosted virtualized services on a data center node; and (ii) flexibility to switch between space-shared and time-shared allocation of processing cores to virtualized services. These compelling features of CloudSim would speed up the development of new resource allocation policies and scheduling algorithms for Cloud computing.

# 2. Key Concepts and Terminologies

This section presents background information on various architectural elements that form the basis for Cloud computing. It also presents requirements of various applications that need to scale across multiple geographically distributed data centers owned by one or more service providers. As development of resource allocation and application scaling techniques and their performance evaluation under various operational scenarios in a real Cloud environment is difficult and hard to repeat; we propose the use of simulation as an alternate approach for achieving the same.

## *2.1 Cloud computing*

Cloud computing can be defined as "a type of parallel and distributed system consisting of a collection of inter-connected and virtualized computers that are dynamically provisioned and presented as one or more unified computing resources based on service-level agreements established through negotiation between the service provider and consumers" [1]. Some examples of emerging Cloud computing infrastructures are Microsoft Azure [2], Amazon EC2, Google App Engine, and Aneka [3].

Emerging Cloud applications such as social networking, gaming portals, business applications, content delivery, and scientific workflows operate at the highest layer of the architecture. Actual usage patterns of many real-world applications vary with time, most of the time in unpredictable ways. These applications have different Quality of Service (QoS) requirements depending on time criticality and users' interaction patterns (online/offline).

## *2.2 Layered Design*

Figure 1 shows the layered design of service-oriented Cloud computing architecture. Physical Cloud resources along with core middleware capabilities form the basis for delivering IaaS. The user-level middleware aims at providing PaaS capabilities. The top layer focuses on application services (SaaS) by making use of services provided by the lower layer services. PaaS/SaaS services are often developed and provided by 3rd party service providers, who are different from IaaS providers [13].

**User-Level Middleware:** This layer includes the software frameworks such as Web 2.0 Interfaces (Ajax, IBM Workplace) that help developers in creating rich, cost-effecting user-interfaces for browser-based applications. The layer also provides the programming environments and composition tools that ease the creation, deployment, and execution of applications in Clouds.

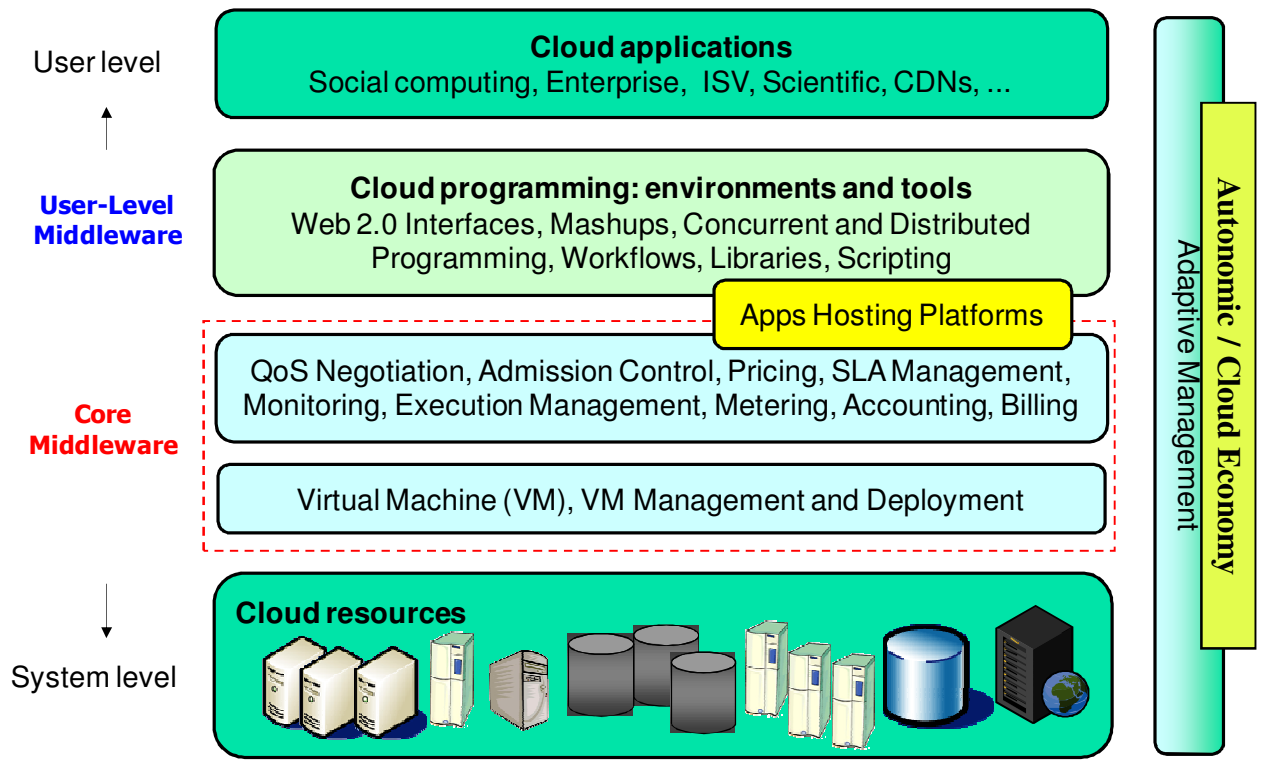

**Figure 1. Layered Cloud Computing Architecture.**

**Core Middleware:** This layer implements the platform level services that provide runtime environment enabling Cloud computing capabilities to application services built using User-Level Middlewares. Core services at this layer includes Dynamic SLA Management, Accounting, Billing, Execution monitoring and management, and Pricing. The well-known examples of services operating at this layer are Amazon EC2, Google App Engine, and Aneka [3].

**System Level:** The computing power in Cloud computing environments is supplied by a collection of data centers,

which are typically installed with hundreds to thousands of servers [9]. At the System Level layer there exist massive physical resources (storage servers and application servers) that power the data centers. These servers are transparently managed by the higher level virtualization [8] services and toolkits that allow sharing of their capacity among virtual instances of servers. These VMs are isolated from each other, which aid in achieving fault tolerant behavior and isolated security context.

## *2.3 Federation (Inter-Networking) of Clouds*

Current Cloud Computing providers have several data centers at different geographical locations over the Internet in order to optimally serve costumers needs around the world. However, existing systems does not support mechanisms and policies for dynamically coordinating load-shredding among different data centers in order to determine optimal location for hosting application services to achieve reasonable service satisfaction levels. Further, the Cloud service providers are unable to predict geographic distribution of users consuming their services, hence the load coordination must happen automatically, and distribution of services must change in response to changes in the load behaviour. Figure 2 depicts such a service-oriented Cloud computing architecture consisting of service consumer's brokering and provider's coordinator services that support utility-driven internetworking of clouds [12]: application scheduling, resource allocation, and workload migration.

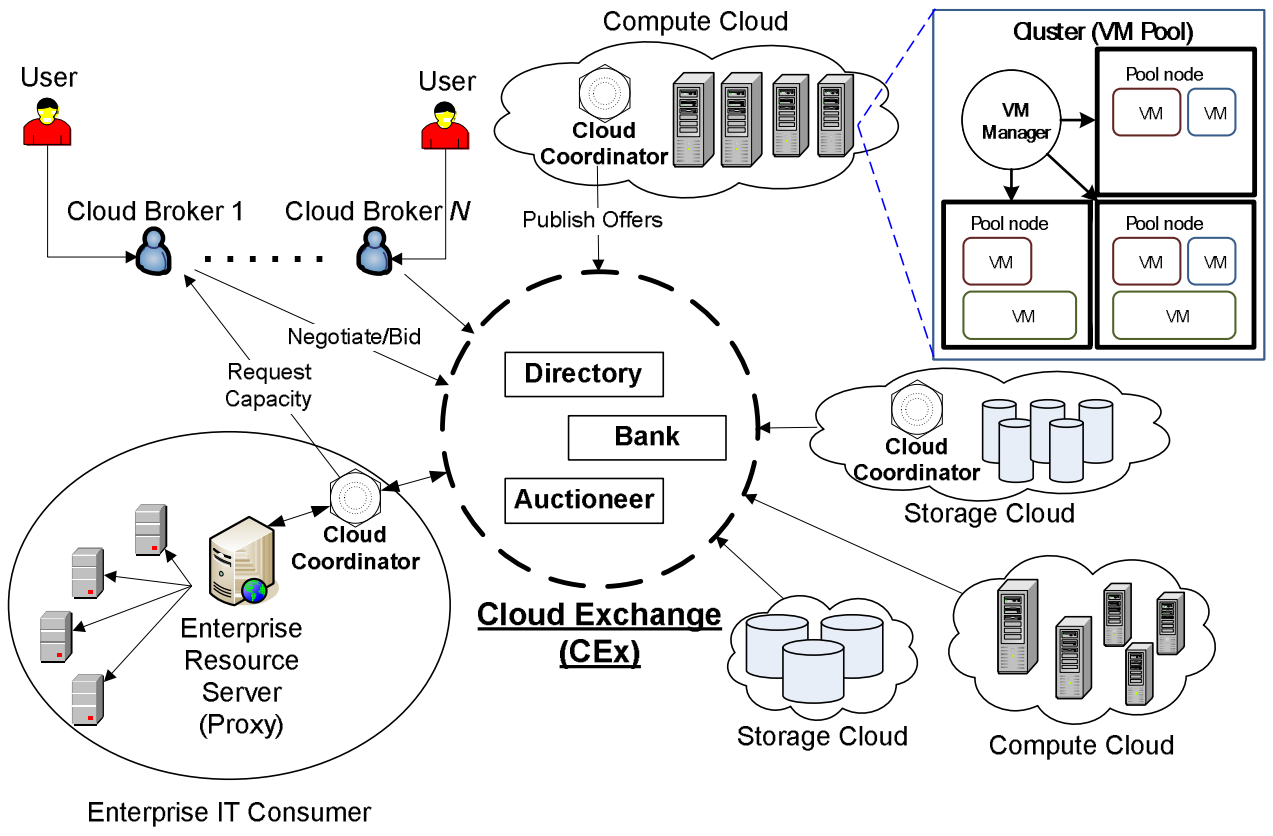

**Figure 2. Clouds and their federated network mediated by a Cloud exchange.**

The Cloud coordinator component is instantiated by each data center that: (i) exports the Cloud services, both infrastructure and platform-level, to the federation; (ii) keeps track of load on the data center and undertakes negotiation with other Cloud providers for dynamic scaling of services across multiple data centers for handling the peak in demands; and (iii) monitors the application execution and oversees that agreed SLAs are delivered. The Cloud brokers acting on behalf of service consumers (users) identify suitable Cloud service providers through the Cloud Exchange and negotiate with Cloud Coordinators for allocation of resources that meets the QoS needs of hosted applications. The Cloud Exchange (CEx) acts as a market maker for bringing together service providers and consumers. It aggregates the infrastructure demands from the Cloud brokers and evaluates them against the available supply currently published by the Cloud Coordinators.

The applications that would benefit from the aforementioned federated Cloud computing system include social networks such as Facebook and MySpace, Content Delivery Networks (CDNs). Social networking sites serve dynamic contents to millions of users, whose access and interaction patterns are difficult to predict. In general, social networking websites are built using multi-tiered web applications such as WebSphere and persistency layers such as the MySQL relational database. Usually, each component will run in a different virtual machine, which can be hosted in data centers owned by different Cloud computing providers. Additionally, each plug-in developer has the freedom to choose which Cloud computing provider offers the services that are more suitable to run his/her plug-in. As a consequence, a typical social networking web application is formed by hundreds of different services, which may be hosted by dozens of Cloud-oriented data centers around the world. Whenever there is a variation in temporal and spatial locality of workload, each application component must dynamically scale to offer good quality of experience to users.

## *2.4 A Case for Simulation and Related Work*

In the past decade, Grids [5] have evolved as the infrastructure for delivering high-performance services for compute and data-intensive scientific applications. To support research and development of new Grid components, policies, and middleware; several Grid simulators, such as GridSim [9], SimGrid [7], and GangSim [4] have been proposed. SimGrid is a generic framework for simulation of distributed applications on Grid platforms. Similarly, GangSim is a Grid simulation toolkit that provides support for modeling of Grid-based virtual organisations and resources. On the other hand, GridSim is an event-driven simulation toolkit for heterogeneous Grid resources. It supports modeling of grid entities, users, machines, and network, including network traffic.

Although the aforementioned toolkits are capable of modeling and simulating the Grid application behaviors (execution, scheduling, allocation, and monitoring) in a distributed environment consisting of multiple Grid organisations, none of these are able to support the infrastructure and application-level requirements arising from Cloud computing paradigm. In particular, there is very little or no support in existing Grid simulation toolkits for modeling of on-demand virtualization enabled resource and application management. Further, Clouds promise to deliver services on subscription-basis in a pay-as-you-go model to Cloud customers. Hence, Cloud infrastructure modeling and simulation toolkits must provide support for economic entities such as Cloud brokers and Cloud exchange for enabling real-time trading of services between customers and providers. Among the currently available simulators discussed in this paper, only GridSim offers support for economic-driven resource management and application scheduling simulation.

Another aspect related to Clouds that should be considered is that research and development in Cloud computing systems, applications and services are in their infancy. There are a number of important issues that need detailed investigation along the Cloud software stack. Topics of interest to Cloud developers include economic strategies for provisioning of virtualized resources to incoming user's requests, scheduling of applications, resources discovery, inter-cloud negotiations, and federation of clouds. To support and accelerate the research related to Cloud computing systems, applications and services; it is important that the necessary software tools are designed and developed to aid researchers.

# 3. CloudSim Architecture

Figure 3 shows the layered implementation of the CloudSim software framework and architectural components. At the lowest layer is the SimJava discrete event simulation engine [6] that implements the core functionalities required for higher-level simulation frameworks such as queuing and processing of events, creation of system components (services, host, data center, broker, virtual machines), communication between components, and management of the simulation clock. Next follows the libraries implementing the GridSim toolkit [9] that support: (i) high level software components for modeling multiple Grid infrastructures, including networks and associated traffic profiles; and (ii) fundamental Grid components such as the resources, data sets, workload traces, and information services.

The CloudSim is implemented at the next level by programmatically extending the core functionalities exposed by the GridSim layer. CloudSim provides novel support for modeling and simulation of virtualized Cloud-based data center environments such as dedicated management interfaces for VMs, memory, storage, and bandwidth. CloudSim layer manages the instantiation and execution of core entities (VMs, hosts, data centers, application) during the simulation period. This layer is capable of concurrently instantiating and transparently managing a large scale Cloud infrastructure consisting of thousands of system components. The fundamental issues such as provisioning of hosts to VMs based on user requests, managing application execution, and dynamic monitoring are handled by this layer. A Cloud provider, who wants to study the efficacy of different policies in allocating its hosts, would need to implement his strategies at this layer by programmatically extending the core VM provisioning functionality. There is a clear distinction at this layer on how a host is allocated to different competing VMs in the Cloud. A Cloud host can be concurrently shared among a number of VMs that execute applications based on user-defined QoS specifications.

The top-most layer in the simulation stack is the User Code that exposes configuration related functionalities for hosts (number of machines, their specification and so on), applications (number of tasks and their requirements), VMs, number of users and their application types, and broker scheduling policies. A Cloud application developer can generate: (i) a mix of user request distributions, application configurations; and (ii) Cloud availability scenarios at this layer and perform robust tests based on the custom configurations already supported within the CloudSim.

As Cloud computing is a rapidly evolving research area, there is a severe lack of defined standards, tools and methods that can efficiently tackle the infrastructure and application level complexities. Hence in the near future there would be a number of research efforts both in academia and industry towards defining core algorithms, policies, application benchmarking based on execution contexts. By extending the basic functionalities already exposed by CloudSim, researchers would be able to perform tests based on specific scenarios and configurations, hence allowing the development of best practices in all the critical aspects related to Cloud Computing.

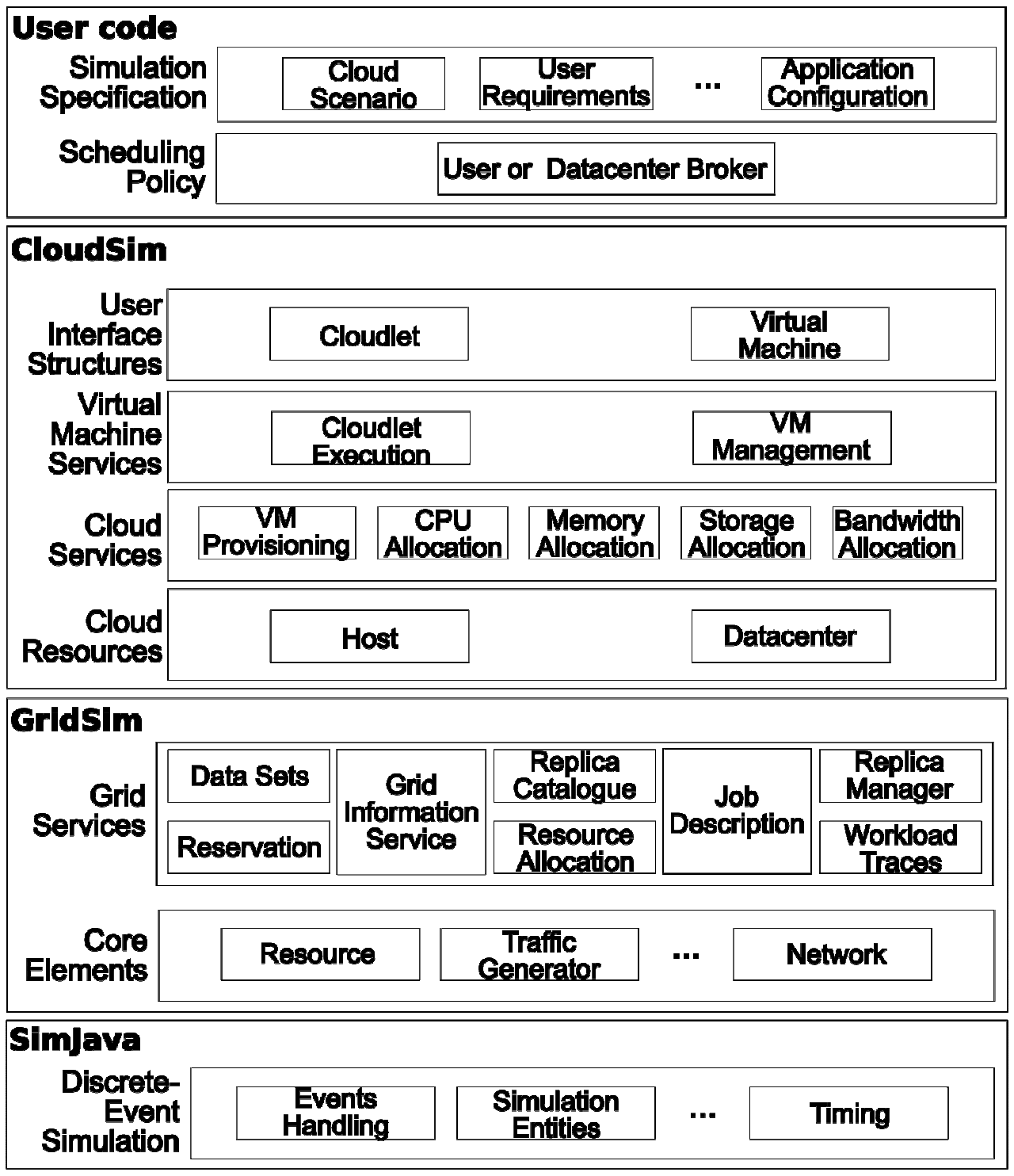

**Figure 3. Layered CloudSim architecture.**

One of the design decisions that we had to make as the CloudSim was being developed was whether to extensively reuse existing simulation libraries and frameworks or not. We decided to take advantage of already implemented and proven libraries such as GridSim and SimJava to handle low-level requirements of the system. For example, by using SimJava, we avoided reimplementation of event handling and message passing among components. This saved us time and cost of software engineering and testing. Similarly, the use of the GridSim framework allowed us to reuse its implementation of networking, information services, files, users, and resources. Since,SimJava and GridSim have been extensively utilized in conducting cutting edge research in Grid resource management by several researchers. Therefore, bugs that may compromise the validity of the simulation have been already detected and fixed. By reusing these long validated frameworks, we were able to focus on critical aspects of the system that are relevant to Cloud computing. At the same time taking advantage of the reliability of components that are not directly related to Clouds.

### 3.1. Modeling the Cloud

The core hardware infrastructure services related to the Clouds are modeled in the simulator by a Datacenter component for handling service requests. These requests are application elements sandboxed within VMs, which need to be allocated a share of processing power on Datacenter's host components. By VM processing, we mean a set of operations related to VM life cycle: provisioning of a host to a VM, VM creation, VM destruction, and VM migration.

A Datacenter is composed by a set of hosts, which are responsible for managing VMs during their life cycles. Host is a component that represents a physical computing node in a Cloud: it is assigned a pre-configured processing capability (expressed in million of instructions per second – MIPS), memory, storage, and a scheduling policy for allocating processing cores to virtual machines. The Host component implements interfaces that support modeling and simulation of both single-core and multi-core nodes.

Allocation of application-specific VMs to Hosts in a Cloud-based data center is the responsibility of the Virtual Machine Provisioner component. This component exposes a number of custom methods for researchers, which aids in implementation of new VM provisioning policies based on optimization goals (user centric, system centric). The default policy implemented by the VM Provisioner is a straightforward policy that allocates a VM to the Host in First-Come-First-Serve (FCFS) basis. The system parameters such as the required number of processing cores, memory and storage as requested by the Cloud user form the basis for such mappings. Other complicated policies can be written by the researchers based on the infrastructure and application demands.

For each Host component, the allocation of processing cores to VMs is done based on a host allocation. The policy takes into account how many processing cores will be delegated to each VM, and how much of the processing core's capacity will effectively be attributed for a given VM. So, it is possible to assign specific CPU cores to specific VMs (a space-shared policy) or to dynamically distribute the capacity of a core among VMs (time-shared policy), and to assign cores to VMs on demand, or to specify other policies.

Each Host component instantiates a VM scheduler component that implements the space-shared or time-shared policies for allocating cores to VMs. Cloud system developers and researchers can extend the VM scheduler component for experimenting with more custom allocation policies. Next, the finer level details related to the time-shared and space-shared policies are described.

### 3.2. Modeling the VM allocation

One of the key aspects that make a Cloud computing infrastructure different from a Grid computing is the massive deployment of virtualization technologies and tools. Hence, as compared to Grids, we have in Clouds an extra layer (the virtualization) that acts as an execution and hosting environment for Cloud-based application services.

Hence, traditional application mapping models that assign individual application elements to computing nodes do not accurately represent the computational abstraction which is commonly associated with the Clouds. For example, consider a physical data center host that has single processing core, and there is a requirement of concurrently instantiating two VMs on that core. Even though in practice there is isolation between behaviors (application execution context) of both VMs, the amount of resources available to each VM is constrained by the total processing power of the host. This critical factor must be considered during the allocation process, to avoid creation of a VM that demands more processing power than the one available in the host, as multiple task units in each virtual machine shares time slices of the same processing core.

To allow simulation of different policies under varying levels of performance isolation, CloudSim supports VM scheduling at two levels: First, at the host level and second, at the VM level. At the host level, it is possible to specify how much of the overall processing power of each core in a host will be assigned to each VM. At the VM level, the VMs assign specific amount of the available processing power to the individual task units that are hosted within its execution engine.

At each level, CloudSim implements the time-shared and space-shared resource allocation policies. To clearly illustrate the difference between these policies and their effect on the application performance, in Figure 4 we show a simple scheduling scenario. In this figure, a host with two CPU cores receives request for hosting two VMs, and each one requiring two cores and running four tasks units: t1, t2, t3 and t4 to be run in VM1, while t5, t6, t7, and t8 to be run in VM2.

Figure 4(a) presents a space-shared policy for both VMs and task units: as each VM requires two cores, only one VM can run at a given instance of time. Therefore, VM2 can only be assigned the core once VM1 finishes the execution of task units. The same happens for tasks hosted within the VM: as each task unit demands only one core, two of them run simultaneously, and the other two are queued until the completion of the earlier task units.

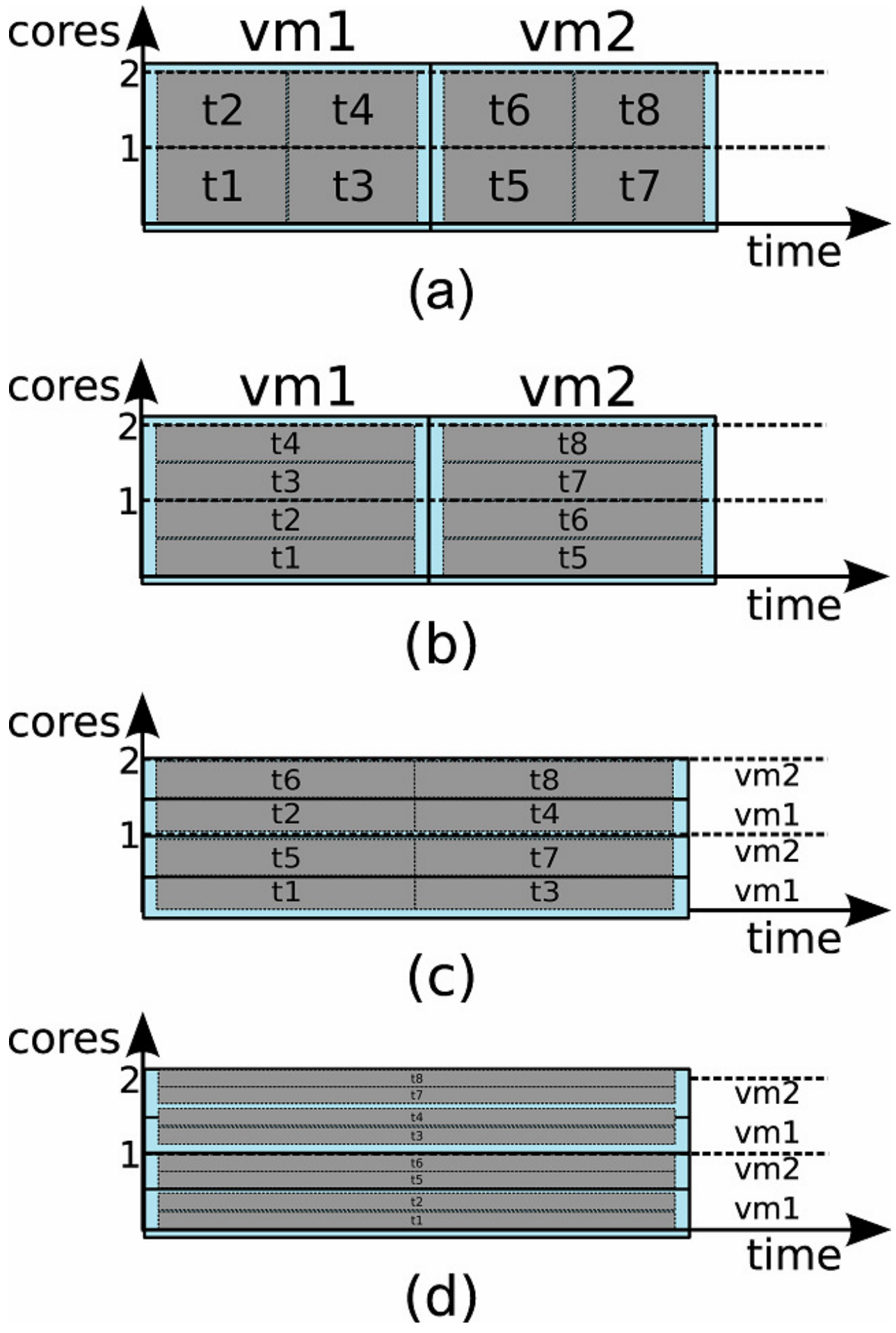

**Figure 4. Effects of different scheduling policies on task execution: (a) Space-shared for VMs and tasks, (b) Space-shared for VMs and time-shared for tasks, (c) Time-shared for VMs, space-shared for tasks, and (d) Time-shared for VMs and tasks.**

In Figure 4(b), a space-shared policy is used for allocating VMs, but a time-shared policy is used for allocating individual task units within VM. Hence, during a VM lifetime, all the tasks assigned to it dynamically context switch until their completion. This allocation policy enables the task units to be scheduled at an earlier time, but significantly affecting the completion time of task units that are ahead the queue.

In Figure 4(c), a time-shared scheduling is used for VMs, and a space-shared one is used for task units. In this case, each VM receives a time slice of each processing core, and then slices are distributed to task units on space-shared basis. As the core is shared, the amount of processing power available to the VM is comparatively lesser than the aforementioned scenarios. As task unit assignment is space-shared, hence only one task can be allocated to each core, while others are queued in for future consideration.

Finally, in Figure 4(d) a time-shared allocation is applied for both VMs and task units. Hence, the processing power is concurrently shared by the VMs and the shares of each VM are concurrently divided among the task units assigned to each VM. In this case, there are no queues either for virtual machines or for task units.

### 3.3. Modeling the Cloud Market

Support for services that act as a market maker enabling capability sharing across Cloud service providers and customer through its match making services is critical to Cloud computing. Further, these services need mechanisms to determine service costs and pricing policies. Modeling of costs and pricing policies is an important aspect to be considered when designing a Cloud simulator. To allow the modeling of the Cloud market, four market-related properties are associated to a data center: cost per processing, cost per unit of memory, cost per unit of storage, and cost per unit of used bandwidth. Cost per memory and storage incur during virtual machine creation. Cost per bandwidth incurs during data transfer. Besides costs for use of memory, storage, and bandwidth, the other cost is associated to use of processing resources. Inherited from the GridSim model, this cost is associated with the execution of user task units. Hence, if VMs were created but no task units were executed on them, only the costs of memory and storage will incur. This behavior may, of course, be changed by users.

## 4. Design and Implementation of CloudSim

The Class design diagram for the simulator is depicted in Figure 5. In this section, we provide finer details related to the fundamental classes of CloudSim, which are building blocks of the simulator.

**DataCenter.** This class models the core infrastructure level services (hardware, software) offered by resource providers in a Cloud computing environment. It encapsulates a set of compute hosts that can be either homogeneous or heterogeneous as regards to their resource configurations (memory, cores, capacity, and storage). Furthermore, every DataCenter component instantiates a generalized resource provisioning component that implements a set of policies for allocating bandwidth, memory, and storage devices.

**DatacenterBroker**. This class models a broker, which is responsible for mediating between users and service providers depending on users' QoS requirements and deploys service tasks across Clouds. The broker acting on behalf of users identifies suitable Cloud service providers through the Cloud Information Service (CIS) and negotiates with them for an allocation of resources that meet QoS needs of users. The researchers and system developers must extend this class for conducting experiments with their custom developed application placement policies.

**SANStorage.** This class models a storage area network that is commonly available to Cloud-based data centers for storing large chunks of data. SANStorage implements a simple interface that can be used to simulate storage and retrieval of any amount of data, at any time subject to the availability of network bandwidth. Accessing files in a SAN at run time incurs additional delays for task unit execution, due to time elapsed for transferring the required data files through the data center internal network.

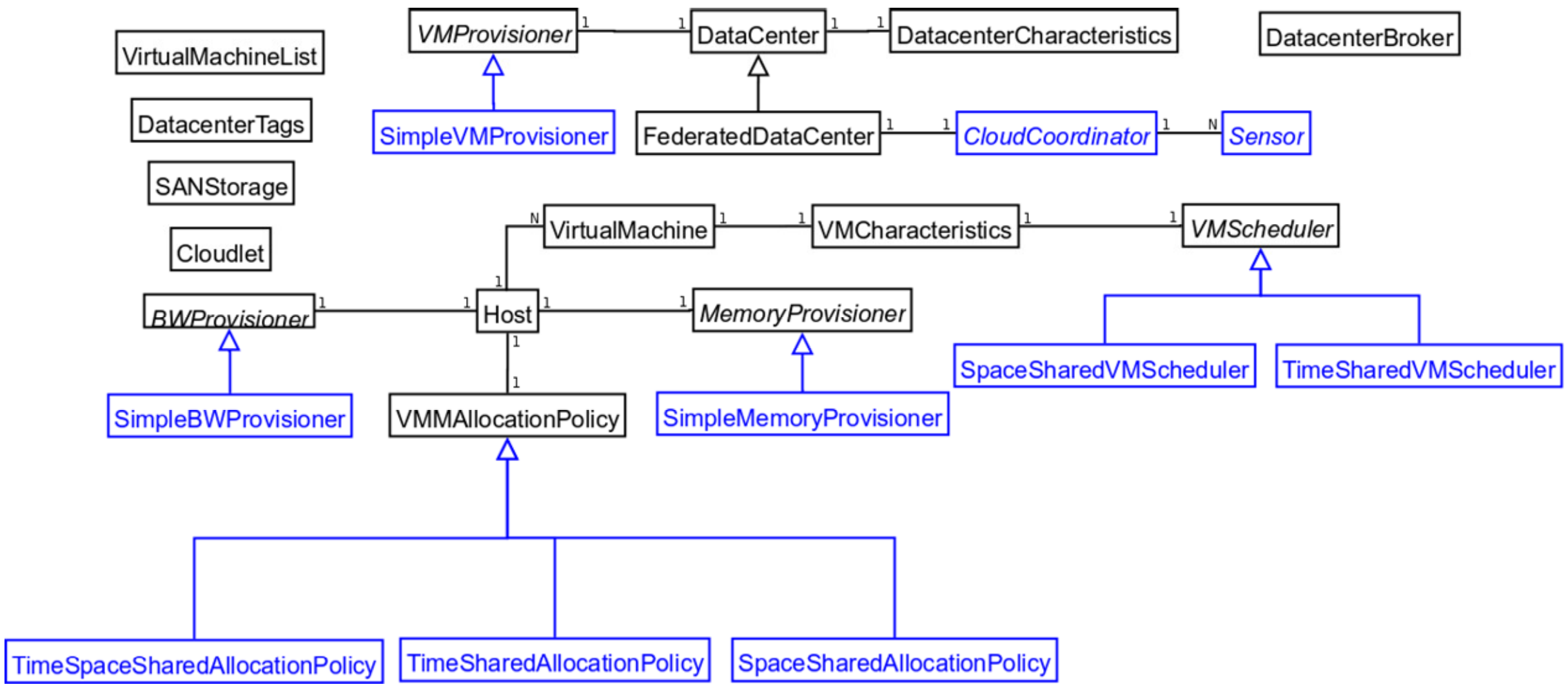

**Figure 5: CloudSim class design diagram.**

**VirtualMachine.** This class models an instance of a VM, whose management during its life cycle is the responsibility of the Host component. As discussed earlier, a host can simultaneously instantiate multiple VMs and allocate cores based on predefined processor sharing policies (space-shared, time-shared). Every VM component has access to a component that stores the characteristics related to a VM, such as memory, processor, storage, and the VM's internal scheduling policy, which is extended from the abstract component called VMScheduling.

**Cloudlet.** This class models the Cloud-based application services (content delivery, social networking, business workflow), which are commonly deployed in the data centers. CloudSim represents the complexity of an application in terms of its computational requirements. Every application component has a pre-assigned instruction length (inherited from GridSim's Gridlet component) and amount of data transfer (both pre and post fetches) that needs to be undertaken for successfully hosting the application.

**CloudCoordinator.** This abstract class provides federation capacity to a data center. This class is responsible for not only communicating with other peer CloudCoordinator services and Cloud Brokers (DataCenterBroker), but also for monitoring the internal state of a data center that plays integral role in load-balancing/application scaling decision making. The monitoring occurs periodically in terms of simulation time. The specific event that triggers the load migration is implemented by CloudSim users through Sensor component. Each sensor may model one specific triggering procedure that may cause the CloudCoordinator to undertake dynamic load-shredding.

**BWProvisioner.** This is an abstract class that models the provisioning policy of bandwidth to VMs that are deployed on a Host component. The function of this component is to undertake the allocation of network bandwidths to set of competing VMs deployed across the data center. Cloud system developers and researchers can extend this class with their own policies (priority, QoS) to reflect the needs of their applications.

**MemoryProvisioner.** This is an abstract class that represents the provisioning policy for allocating memory to VMs. This component models policies for allocating physical memory spaces to the competing VMs. The execution and deployment of VM on a host is feasible only if the MemoryProvisioner component determines that the host has the amount of free memory, which is requested for the new VM deployment.

**VMProvisioner.** This abstract class represents the provisioning policy that a VM Monitor utilizes for allocating VMs to Hosts. The chief functionality of the VMProvisioner is to select available host in a data center, which meets the memory, storage, and availability requirement for a VM deployment. The default SimpleVMProvisioner implementation provided with the CloudSim package allocates VMs to the first available Host that meets the aforementioned requirements. Hosts are considered for mapping in a sequential order. However, more complicated policies can be easily implemented within this component for achieving optimized allocations, for example, selection of hosts based on their ability to meet QoS requirements such as response time, budget.

**VMMAllocationPolicy.** This is an abstract class implemented by a Host component that models the policies (space-shared, time-shared) required for allocating processing power to VMs. The functionalities of this class can easily be overridden to accommodate application specific processor sharing policies.

### 4.1. Entities and threading

As the CloudSim programmatically builds upon the SimJava discrete event simulation engine, it preserves the SimJava's threading model for creation of simulation entities. A programming component is referred to as an entity if it directly extends the core Sim_Entity component of SimJava, which implements the Runnable interface. Every entity is capable of sending and receiving messages through the SimJava's shared event queue. The message propagation (sending and receiving) occurs through input and output ports that SimJava associates with each entity in the simulation system. Since threads incur a lot of memory and processor context switching overhead, having a large number of threads/entities in a simulation environment can be performance bottleneck due to limited scalability. To counter this behavior, CloudSim minimizes the number of entities in the system by implementing only the core components (Users and Datacenters) as the inherited members of SimJava entities. This design decision is significant as it helps CloudSim in modeling a really large scale simulation environment on a computing machine (desktops, laptops) with moderate processing capacity. Other key CloudSim components such as VMs, provisioning policies, hosts are instantiated as standalone objects, which are lightweight and do not compete for processing power.

Hence, regardless of the number of hosts in a simulated data center, the runtime environment (Java virtual machine) needs to manage only two threads (Datacenter and Broker). As the processing of task units is handled by respective VMs, therefore their (task) progress must be updated and monitored after every simulation step. To handle this, an internal event is generated regarding the expected completion time of a task unit to inform the Datacenter entity about the future completion events. Thus, at each simulation step, each Datacenter invokes a method called updateVMsProcessing() for every host in the system, to update processing of tasks running within the VMs. The argument of this method is the current simulation time and the return type is the next expected completion time of a task running in one of the VMs on a particular host. The least time among all the finish times returned by the hosts is noted for the next internal event.

At the host level, invocation of updateVMsProcessing() triggers an updateGridletsProcessing() method, which directs every VM to update its tasks unit status (finish, suspended, executing) with the Datacenter entity. This method implements the similar logic as described previously for updateVMsProcessing() but at the VM level. Once this method is called, VMs return the next expected completion time of the task units currently managed by them. The least completion time among all the computed values is send to the Datacenter entity. As a result, completion times are kept in a queue that is queried by Datacenter after each event processing step. If there are completed tasks waiting in the queue, then they are removed from it and sent back to the user.

### 4.2. Communication among Entities

Figure 6 depicts the flow of communication among core CloudSim entities. In the beginning of the simulation, each Datacenter entity registers itself with the CIS (Cloud Information Service) Registry. CIS provides database level match-making services for mapping user requests to suitable Cloud providers. Brokers acting on behalf of users consult the CIS service about the list of Clouds who offer infrastructure services matching user's application requirements. In case the match occurs the broker deploys the application with the Cloud that was suggested by the CIS.

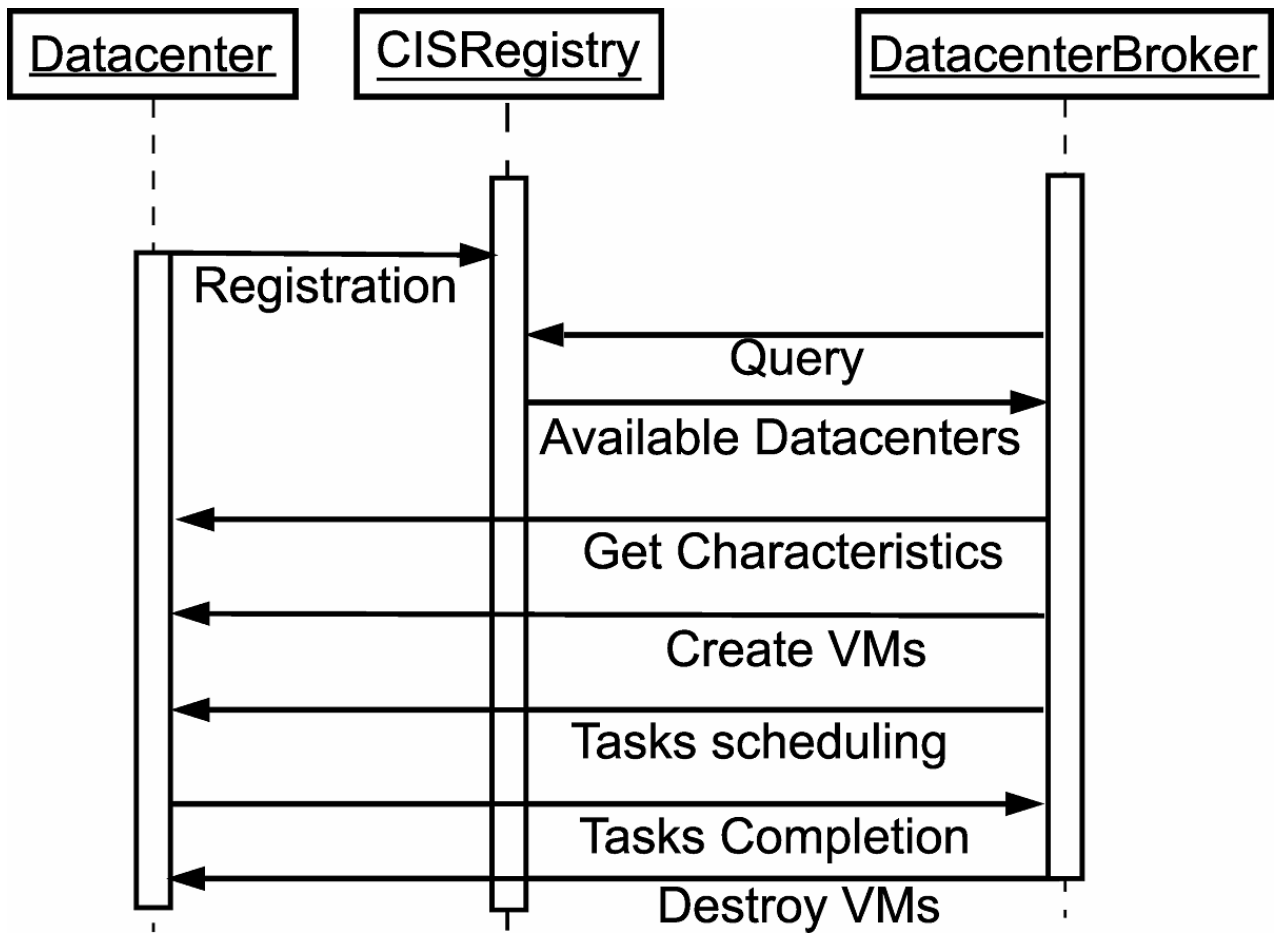

**Figure 6. Simulation data flow.**

The communication flow described so far relates to the basic flow in a simulated experiment. Some variations in this flow are possible depending on policies. For example, messages from Brokers to Datacenters may require a confirmation, from the part of the Datacenter, about the execution of the action, or the maximum number of VMs a user can create may be negotiated before VM creation.

## 5. Experiments and Evaluation

In this section, we present experiments and evaluation that we undertook in order to quantify the efficiency of CloudSim in modeling and simulating Cloud computing environments. The experiments were conducted on a Celeron machine having configuration: 1.86GHz with 1MB of L2 cache and 1 GB of RAM running a standard Ubuntu Linux version 8.04 and JDK 1.6.

To evaluate the overhead in building a simulated Cloud computing environment that consists of a single data center, a broker and a user, we performed series of experiments. The number of hosts in the data center in each experiment was varied from 100 to 100000. As the goal of these tests were to evaluate the computing power requirement to instantiate the Cloud simulation infrastructure, no attention was given to the user workload. For the memory test, we profile the total physical memory used by the hosting computer in order to fully instantiate and load the CloudSim environment. The total delay in instantiating the simulation environment is the time difference between the following events: (i) the time at which the runtime environment (Java virtual machine) is directed to load the CloudSim program; and (ii) the instance at which CloudSim's entities and components are fully initialized and are ready to process events.

Figures 7 and 8 present, respectively, the amount of time and the amount of memory is required to instantiate the experiment when the number of hosts in a data center increases. The growth in memory consumption (see Fig. 8) is linear, with an experiment with 100000 machines demanding 75MB of RAM. It makes our simulation suitable to run even on simple desktop computers with moderated processing power because CloudSim memory requirements, even for larger simulated environments can easily be provided by such computers.

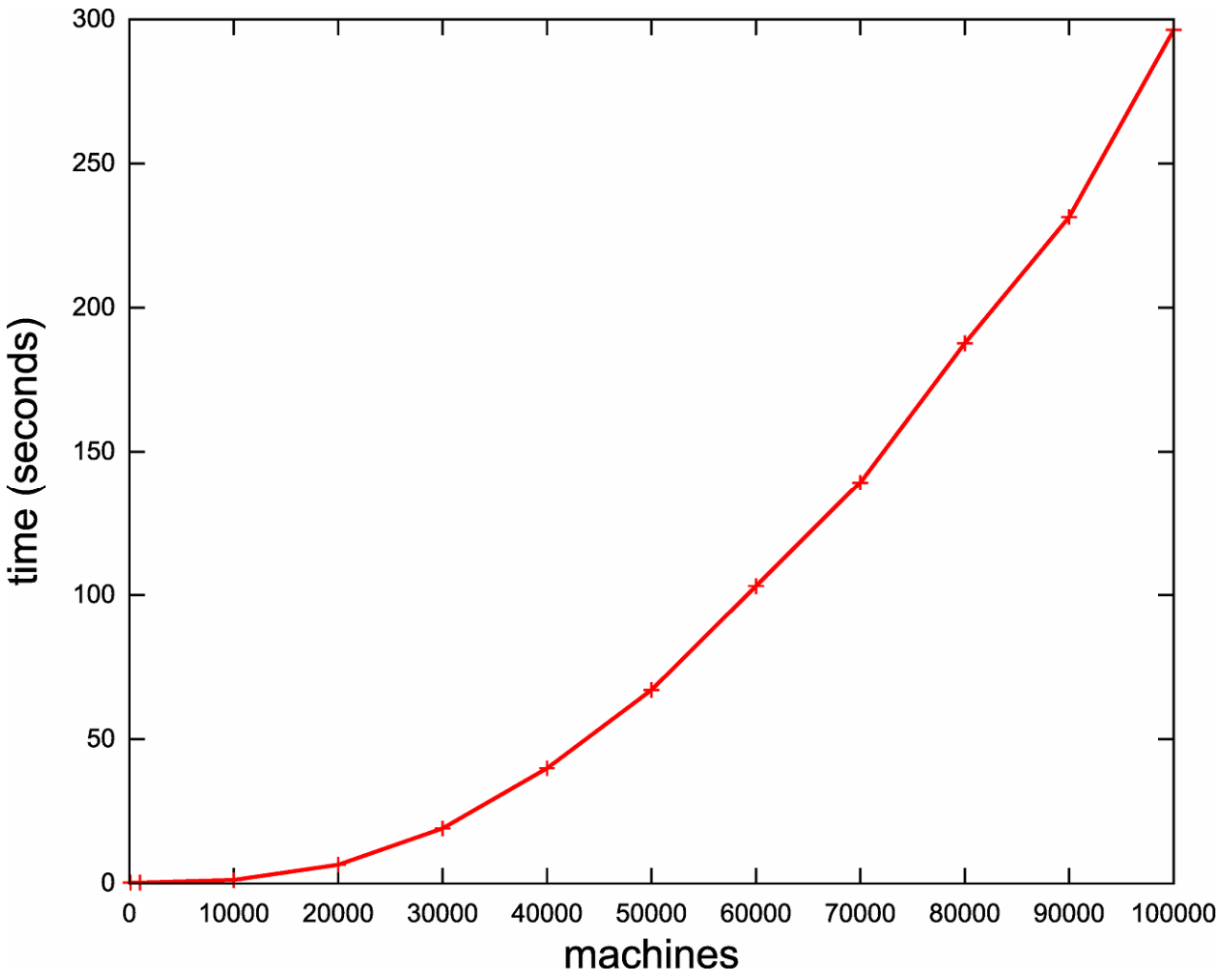

**Figure 7. Time to simulation instantiation.**

Regarding time overhead related to simulation instantiation, the growth in terms of time increases exponentially with the number of hosts/machines. Nevertheless, the time to instantiate 100000 machines is below 5 minutes, which is reasonable considering the scale of the experiment. Currently, we are investigating the cause of this behavior to avoid it in future versions of CloudSim.

The next test aimed at quantifying the performance of CloudSim's core components when subjected to user workloads such as VM creation, task unit execution. The simulation environment consisted of a data center with 10000 hosts, where each host was modeled to have a single CPU core (1000MIPS), 1GB of RAM memory and 2TB of storage. Scheduling policy for VMs was Space-shared, which meant only one VM was allowed to be hosted in a host at a given instance of time. We modeled the user (through the DatacenterBroker) to request creation of 50 VMs having following constraints: 512MB of physical memory, 1 CPU core and 1GB of storage. The application unit was modeled to consist of 500 task units, with each task unit requiring 1200000 million instructions (20 minutes in the simulated hosts) to be executed on a host. As networking was not a concern in these experiments, task units required only 300kB of data to be transferred to and from the data center.

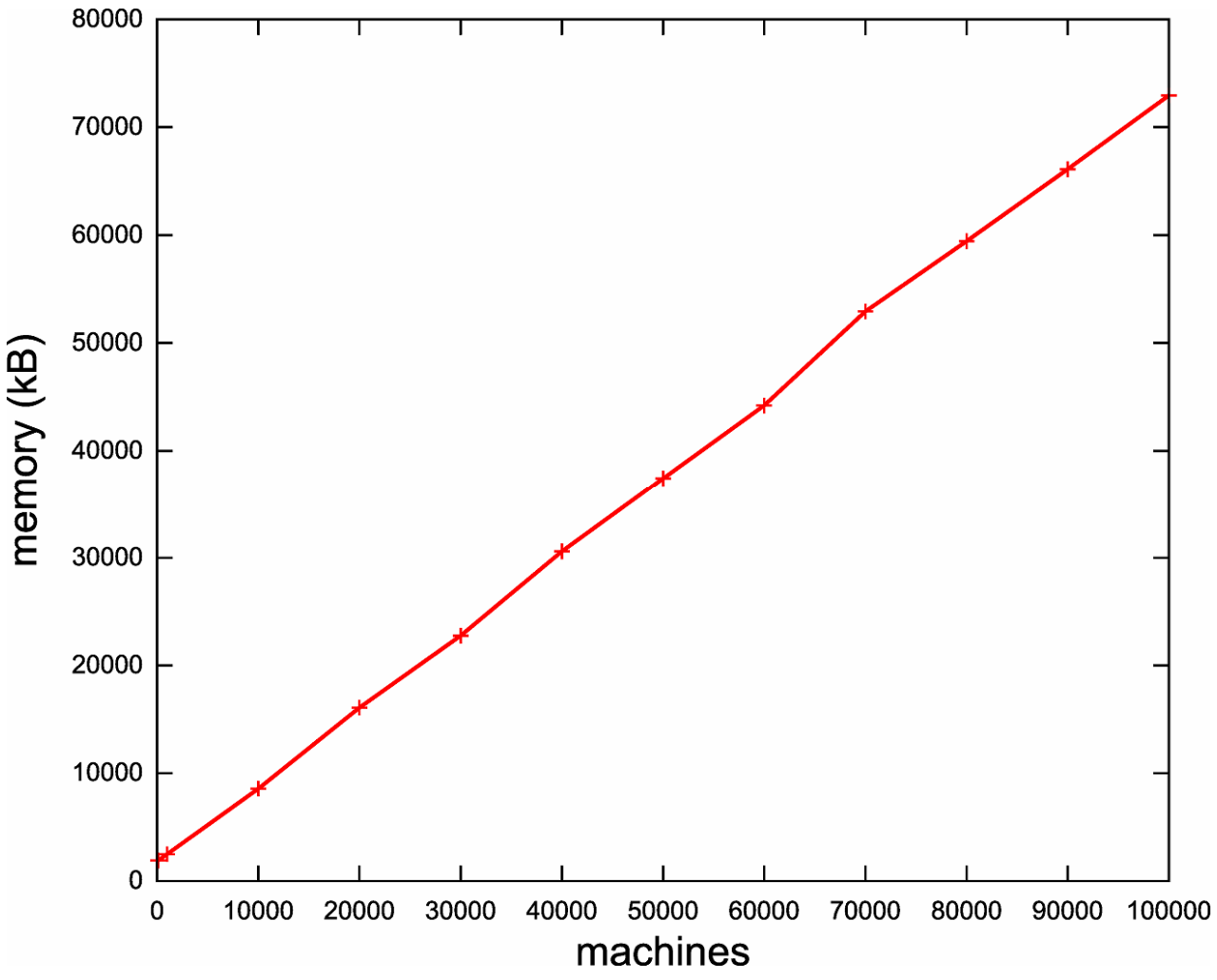

**Figure 8. Memory usage in resources instantiation.**

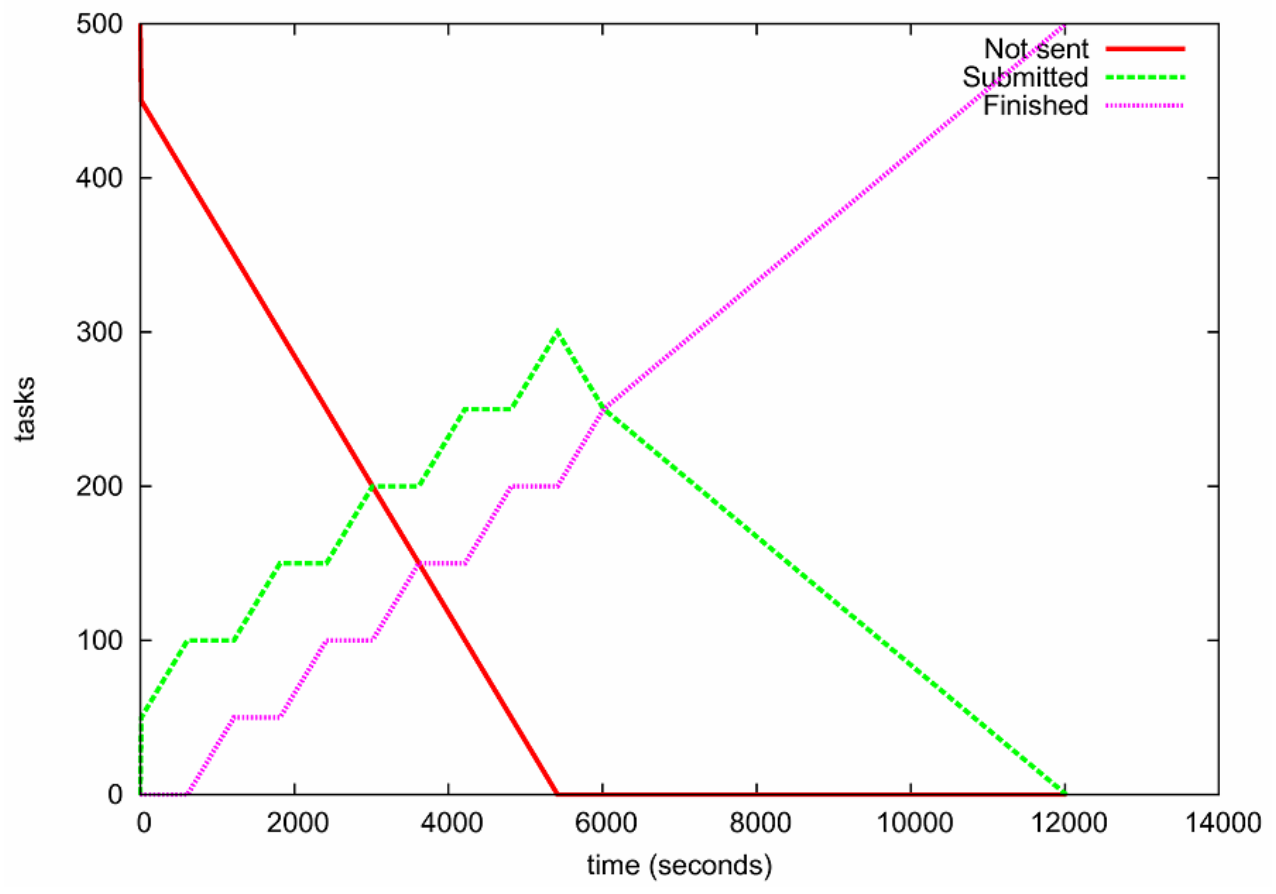

**Figure 9. Tasks execution with space-shared scheduling of tasks.**

After creation of VMs, task units were submitted in groups of 50 (one submitted to each VM) every 10 minutes. The VM were configured to use both space-shared and time-shared policies for allocating tasks units to the processing cores.

Figures 9 and 10 present task units progress status with increase in simulation steps (time) for the space-shared test and for the time-shared tests respectively. As expected, in the space-shared case every task took 20 minutes for completion as they had dedicated access to the processing core. Since, in this policy each task unit had its own dedicated core, the number of incoming tasks or queue size did not affect execution time of individual task units.

However, in the time-shared case execution time of each task varied with increase in number of submitted taks units. Using this policy, execution time is significantly affected as the processing core is concurrently context switched among the list of scheduled tasks. The first group of 50 tasks was able to complete earlier than the other ones because in this case the hosts were not over-loaded at the beginning of execution. To the end, as more tasks reached completion, comparatively more hosts became available for allocation. Due to this we observed improved response time for the tasks as shown in Figure 10.

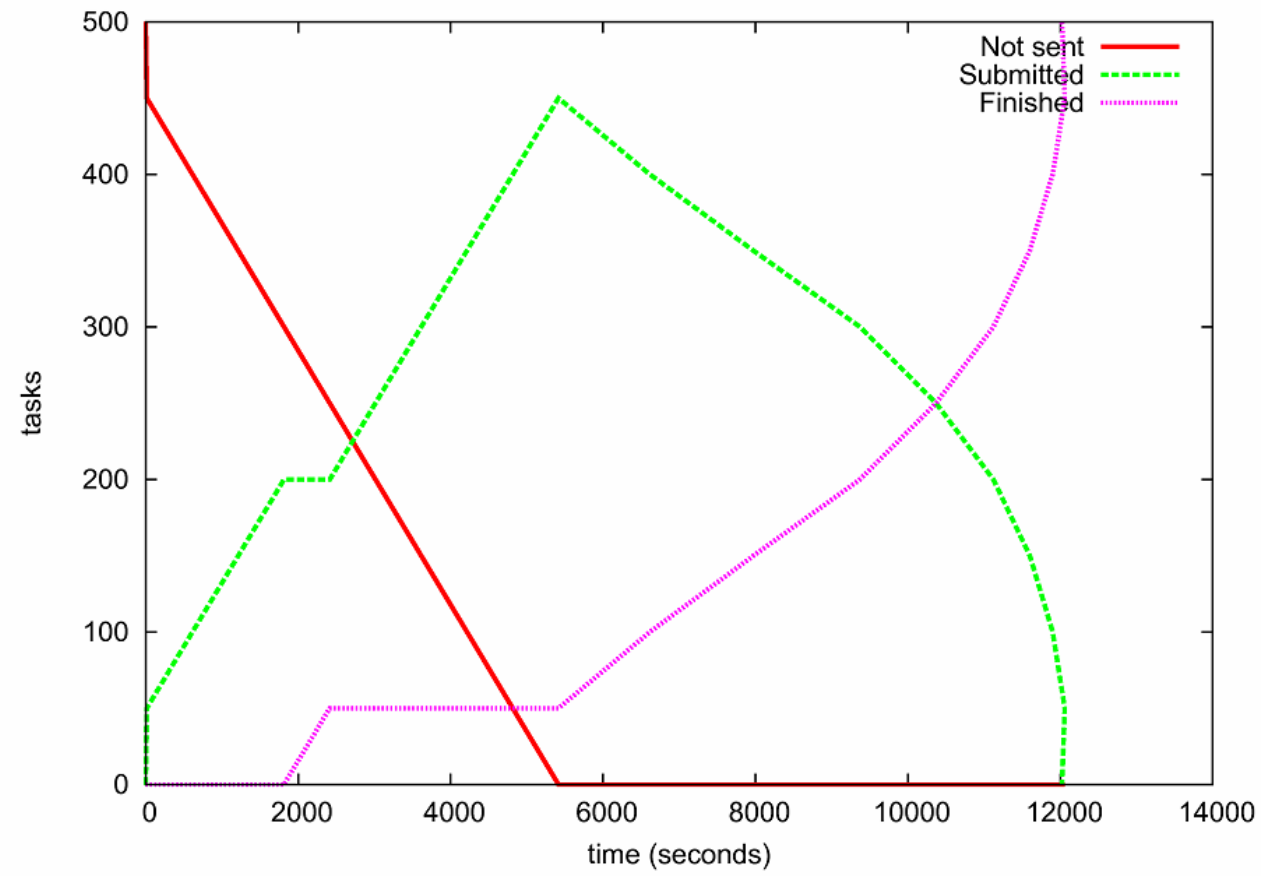

**Figure 10. Task execution with time-shared scheduling of tasks.**

**Evaluating Federated Cloud Computing Components**

This experiment is aimed at testing CloudSim components that form the basis for simulating federated Cloud computing environments. To this end, a simulation environment that models federation of 3 data centers and a user are created. Every data center instantiates a sensor component, which is responsible for dynamically sensing the availability information related to the local hosts. Next, the sensed statistics are reported to the Cloud Coordinator that utilizes the information in undertaking load-migration decisions. We evaluate a straightforward load-migration policy that performs online migration of VMs among federated data centers only if the origin data center does not have the requested number of free VM slots available. The migration process involves the following steps: (i) creating a virtual machine instance that has the same configuration, which is supported at the destination data center; and (ii) migrating the Cloudlets assigned to the original virtual machine to the newly instantiated virtual machine at the destination data center. The federated network of data centers is created based on the topology shown in Figure 11.

Every data center in the system is modeled to have 50 computing hosts, 10GB of memory, 2TB of storage, 1 processor with 1000 MIPS of capacity, and a time-shared VM scheduler. Data center broker on behalf of the user requests instantiation of a VM that requires 256MB of memory, 1GB of storage, 1 CPU, and time-shared Cloudlet scheduler. The broker requests instantiation of 25 VMs and associates one Cloudlet to each VM to be executed. These requests are originally submitted with the Datacenter 0. Each Cloudlet is modeled to be having 1800000 MIs. The simulation experiments were run under the following system configurations: (i) first a federated network of clouds is available, hence data centers are able to cope with peak in demands by migrating the excess of load to the least loaded ones; and (ii) second, the data centers are modeled as independent entities (without federation). All the workload submitted to a data center must be processed and executed locally.

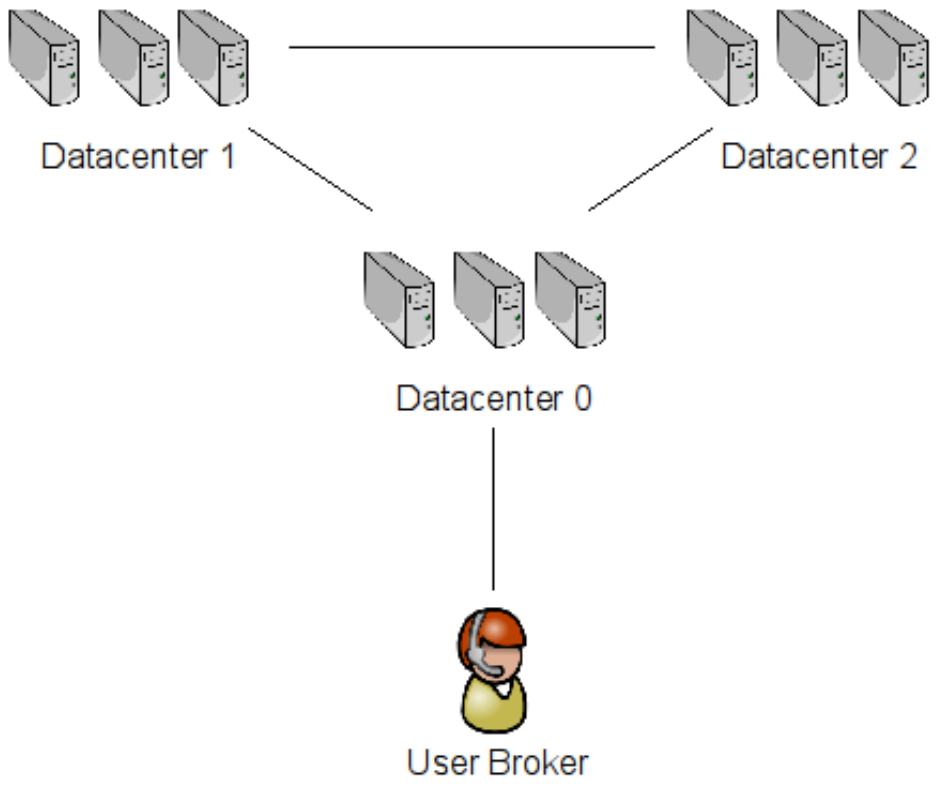

**Figure 11: A network topology of federated Data Centers.**

Table 1 shows the average turn-around time for each Cloudlet and the overall makespan of the user application for both cases. A user application consists of one or more Cloudlets with sequential dependencies. The simulation results reveal that the availability of federated infrastructure of clouds reduces the average turn-around time by more than 50%, while improving the makespan by 20%. It shows that, even for a very simple load-migration policy, availability of federation brings significant benefits to user's application performance.

**Table 1: Performance Results.**

| Performance Metrics | With Federation | Without Federation |
|---|---|---|
| Average Turn Around Time (Secs) | 2221.13 | 4700.1 |
| Makespan (Secs) | 6613.1 | 8405 |

## 6. Conclusion and Future Work

The recent efforts to design and develop Cloud technologies focus on defining novel methods, policies and mechanisms for efficiently managing Cloud infrastructures. To test these newly developed methods and policies, researchers need tools that allow them to evaluate the hypothesis prior to real deployment in an environment where one can reproduce tests. Simulation-based approaches in evaluating Cloud computing systems and application behaviors offer significant benefits, as they allow Cloud developers: (i) to test performance of their provisioning and service delivery policies in a repeatable and controllable environment free of cost; and (ii) to tune the performance bottlenecks before real-world deployment on commercial Clouds.

To meet these requirements, we developed the CloudSim toolkit for modeling and simulation of extensible Clouds. As a completely customizable tool, it allows extension and definition of policies in all the components of the software stack, which makes it suitable as a research tool that can handle the complexities arising from simulated environments. As future work, we are planning to incorporate new pricing and provisioning policies to CloudSim, in order to offer a built-in support to simulate the currently available Clouds. Modeling and simulation of such environments that consist of providers encompassing multiple services and routing boundaries present unique challenges. They include providing support for practical and concrete network models that capture the message routing and latency behavior ambient on the Internet. To address this, we intend to extend CloudSim by implementing the BRITE topology model for networking multiple Clouds.

Further, recent studies have revealed that data centers consume unprecedented amount of electrical power, hence they incur massive capital expenditure for day-to-day operation and management. For example, a Google data center consumes power as much as a city such as San Francisco. The socio-economic factors and environmental conditions of the geographical region, where a data center is hosted directly influences total power bills incurred. For instance, a data center hosted in a location where power cost is low and has less hostile weather conditions, would incur comparatively lesser expenditure in power bills. To achieve simulation of the aforementioned Cloud computing environments, much of our future work would investigate new models and techniques for allocation of services to applications depending on energy efficiency and expenditure of service providers.

## Acknowledgements

This work is partially supported by the Australian Department of Innovation, Industry, Science and Research (DIISR) and the Australian Research Council (ARC) through the International Science Linkage and the Discovery Projects programs respectively. We would like to thank Marcos Assunção for proof reading the paper.

## References

[1] R. Buyya, C. S. Yeo, and S. Venugopal. Market-oriented cloud computing: Vision, hype, and reality for delivering IT services as computing utilities. *Proceedings of the 10th IEEE International Conference on High Performance Computing and Communications*, 2008.

[2] D. Chappell. *Introducing the Azure services platform*. White paper, Oct. 2008.

[3] X. Chu et al. Aneka: Next-generation enterprise grid platform for e-science and e-business applications. *Proceedings of the 3rd IEEE International Conference on e-Science and Grid Computing*, 2007.

[4] C. L. Dumitrescu and I. Foster. GangSim: a simulator for grid scheduling studies. *Proceedings of the IEEE International Symposium on Cluster Computing and the Grid,* 2005.

[5] I. Foster and C. Kesselman (editors). *The Grid: Blueprint for a New Computing Infrastructure*. Morgan Kaufmann, 1999.

[6] F. Howell and R. Mcnab. SimJava: A discrete event simulation library for java. *Proceedings of the first International Conference on Web-Based Modeling and Simulation,* 1998.

[7] A. Legrand, L. Marchal, and H. Casanova. Scheduling distributed applications: the SimGrid simulation framework. *Proceedings of the 3rd IEEE/ACM International Symposium on Cluster Computing and the Grid*, 2003.

[8] J. E. Smith and R. Nair. *Virtual Machines: Versatile platforms for systems and processes.* Morgan Kauffmann, 2005.

[9] R. Buyya and M. Murshed. GridSim: A Toolkit for the Modeling and Simulation of Distributed Resource Management and Scheduling for Grid Computing. *Concurrency and Computation: Practice and Experience*, 14(13-15), Wiley Press, Nov.-Dec., 2002.

[10] A. Weiss. Computing in the clouds. *NetWorker*, 11(4):16–25, Dec. 2007.

[11] M. Armbrust, A. Fox, R. Griffith, A. Joseph, R. Katz, A. Konwinski, G. Lee, D. Patterson, A. Rabkin, I. Stoica, M. Zaharia. *Above the Clouds: A Berkeley View of Cloud computing*. Technical Report No. UCB/EECS-2009-28, University of California at Berkley, USA, Feb. 10, 2009.

[12] R. Ranjan and R. Buyya. Decentralized Overlay for Federation of Enterprise Clouds. *Handbook of Research on Scalable Computing Technologies*, K. Li et. al. (ed), IGI Global, USA, 2009 (in press).

[13] R. Buyya, C. S. Yeo, S. Venugopal, J. Broberg, and I. Brandic. Cloud Computing and Emerging IT Platforms: Vision, Hype, and Reality for Delivering Computing as the 5th Utility. *Future Generation Computer Systems*, 25(6): 599-616, Elsevier Science, Amsterdam, The Netherlands, June 2009.